\documentclass[preprint2]{aastex}
\usepackage{color}
\usepackage{txfonts}
\usepackage{breqn}
\usepackage{psfrag,graphicx,natbib}
\shorttitle{The formation of the first galaxies}

\shortauthors{Latif et al.}

\def\Rev. Mod. Phys{Rev.~Mod.~Phys}

\begin{document}

\title {The formation of massive primordial stars in the presence of moderate UV backgrounds}

\author{M.~A.~Latif \altaffilmark{1}, D.~R.~G.~Schleicher \altaffilmark{1},  S. Bovino \altaffilmark{1}, T. Grassi \altaffilmark{2,3}, M. Spaans\altaffilmark{4} }
\affil{Institut f\"ur Astrophysik, Georg-August-Universit\"at, Friedrich-Hund-Platz 1, 37077 G\"ottingen, Germany}
\affil{Centre for Star and Planet Formation, Natural History Museum of Denmark, \O ster Voldgade 5-7, DK-1350 Copenhagen, Denmark}
\affil{Niels Bohr Institute, University of Copenhagen, Juliane Maries Vej 30, DK-2100 Copenhagen, Denmark}
\affil{Kapteyn Astronomical Institute, University of Groningen, The Netherlands}
\email{mlatif@astro.physik.uni-goettingen.de}
\newcommand{\ch}[1]{\textcolor{red}{\textbf{#1}}}
\date{}

\bibliographystyle{apj}

\begin{abstract}
{
Radiative feedback from populations II stars played a vital role in early structure formation. Particularly, photons below the Lyman limit can escape the star forming regions and produce a background ultraviolet (UV) flux which consequently may influence the pristine halos far away from the radiation sources. These photons can quench the formation of molecular hydrogen by photo-detachment of $\rm H^{-}$. In this study, we explore the impact of such UV radiation on fragmentation in massive primordial halos of a few times $\rm 10^{7}$~M${_\odot}$. To accomplish this goal, we perform high resolution cosmological simulations for two distinct halos and vary the strength of the impinging background UV field in units of $\rm J_{21}$. We further make use of sink particles to follow the evolution for 10,000 years after reaching the maximum refinement level. No vigorous fragmentation is observed in UV illuminated halos while the accretion rate changes according to the thermal properties. Our findings show that a few 100-10, 000 solar mass protostars are formed when halos are irradiated by $\rm J_{21}=10-500$ at $\rm z>10$ and suggest a strong relation between the strength of UV flux and mass of a protostar. This mode of star formation is quite different from minihalos, as higher accretion rates of about $\rm 0.01-0.1$ M$_{\odot}$/yr are observed by the end of our simulations. The resulting massive stars are the potential cradles for the formation of intermediate mass black holes at earlier cosmic times and contribute to the formation of a global X-ray background.   
}

\end{abstract}



\section{Introduction}
The first generation of stars so-called population III stars ushered the Universe out of the cosmic dark ages and brought the first light in the cosmos. They are presumed to be assembled in dark matter halos of $\rm 10^{5}-10^{6}~M_{\odot}$ at $\rm z=20-30$ \citep{2002Sci...295...93A,2011MNRAS.413.1184J,Clark11,Greif12,2012MNRAS.422..290S,2013ApJ...772L...3L,2014MNRAS.441.2181B} and influenced the subsequent structure formation via mechanical, chemical and radiative feedback \citep{2005SSRv..116..625C,2008A&A...490..521S,2012A&A...540A.101L,2011MNRAS.414.1145M}. Pop III stars enriched the intergalactic medium with metals and led to the second generation of stars known as Pop II stars.  According to the hierarchical paradigm of structure formation,  the first galaxies are formed in massive primordial halos of $\rm 10^{7}-10^{8}~M_{\odot}$  at  about $\rm z=15$ most likely hosting both stellar populations \citep{2008MNRAS.387.1021G, 2008ApJ...682..745W,2009Natur.459...49B,2011A&A...532A..66L}.

During the epoch of reionization, radiation emitted by the Pop II stars not only photo-ionized the gas but also photo-dissociated $\rm H_2$ and HD molecules. The molecular hydrogen is the only coolant in primordial gas which can bring the gas temperature down to few hundred K and may induce star formation. In the presence of an intense UV flux, the formation of $\rm H_{2}$ remains suppressed (due to photo-detachment of $\rm H^{-}$ which is the main pathway of $\rm H_{2}$ formation) in pristine halos. Under these conditions, cooling only proceeds via Lyman alpha radiation in massive primordial halos of $\rm 10^{7}-10^{8}~M_{\odot}$ and massive objects are expected to form and are known as direct collapse black holes (DCBHs) \citep{2002ApJ...569..558O,2006ApJ...652..902S,2006MNRAS.370..289B,2006MNRAS.371.1813L,2010A&ARv..18..279V,2010ApJ...712L..69S,2011MNRAS.411.1659L,2011MNRAS.410..919J,2012RPPh...75l4901V,2012arXiv1203.6075H,2013arXiv1301.5567P,2013MNRAS.433.1607L,2013ApJ...774...64W,2013ApJ...771...50A,2013MNRAS.436.2989L,2014MNRAS.tmp..537Y,2014MNRAS.440.2969L,2014arXiv1404.4630I}. 

The formation of DCBHs requires a critical strength of UV flux above which halos remain $\rm H_{2}$ free \citep{2001ApJ...546..635O,2010MNRAS.402.1249S,2014arXiv1404.5773L,2014arXiv1405.2081J}. Such a UV flux can only exist in the close vicinity of star forming galaxies \citep{2008MNRAS.391.1961D,2012MNRAS.425.2854A,2014arXiv1403.5267A,2014arXiv1405.6743D} and the mass scales of the resulting stars have been explored in recent studies. In fact, high resolution numerical simulations show that supermassive stars of  $\rm \sim10^{5}~M_{\odot}$ form in the presence of strong UV flux \citep{2003ApJ...596...34B,2008ApJ...682..745W,2009MNRAS.393..858R,2011MNRAS.411.1659L,2013MNRAS.432..668L,2013MNRAS.430..588L,2013MNRAS.433.1607L,2013MNRAS.436.2989L,2014MNRAS.439.1160R} and agree with theoretical predictions \citep{2008MNRAS.387.1649B,2010MNRAS.402..673B,2011MNRAS.414.2751B,2012ApJ...756...93H,2012MNRAS.421.2713B,2013A&A...558A..59S,2013ApJ...778..178H,2013ApJ...768..195W}. Employing a cosmological framework following the formation and accretion of the resulting supermassive objects, Ferrara et al. (2014) have derived detailed predictions on the mass function of the first high-mass black holes.

On the other hand, moderate strengths of the background UV flux may occur more often during the epoch of reionization without requiring the presence of nearby sources. In the ubiquity of such moderate UV fluxes, the formation of $\rm H_{2}$ does occur and may lead to star formation. Particularly, during the epoch of reionization the dominant background flux is emitted by Pop II stars. The strength of such UV flux to dissociate $\rm H_{2}$ formation was quantified in a recent study by \cite{2014arXiv1404.5773L} and found that $\rm J_{21}^{crit}$ ranges from 400-700 higher than previous estimates. Therefore, halos exposed below $\rm J_{21}^{crit}$ are expected to be more abundant, for $\rm J_{21} =10$ the fraction of halos is seven orders of magnitude higher compared to the $\rm J_{21} =500$ \cite[see figure C1 of][]{2014arXiv1405.6743D}. Thus it is desirable to explore the typical mass scales of stars forming in halos illuminated by moderate UV fluxes emitted by Pop II stars. \cite{2012MNRAS.426.1159S} have performed three dimensional simulations to study the fragmentation in massive halos irradiated by Lyman-Werner flux of strength $\rm J_{21}=100$ for $\rm T_{*}=10^5~K$ and found that a dense turbulent core of 10$^{4}$~M$_{\odot}$ forms in the center of the halo. 

In this study, we explore the mass scales of stars formed under the moderate strengths of the background UV flux. To achieve this goal, we perform three dimensional cosmological simulations for two halos of a few times $\rm 10^{7}~M_{\odot}$ and vary the strength of UV flux (hereafter  called $\rm J_{21}$, i.e. UV flux with energy below 13.6 eV). We employed a Jeans resolution of 32 cells during the course of simulations and make use of sink particles to follow the evolution for 10, 000 years after reaching the maximum refinement level in simulations. The main objective of this work is to determine the mass scale of stars for a moderate UV radiation field. This has potential implications for the formation of intermediate mass black holes in stellar clusters at earlier cosmic times. We note that the expected ionizing UV feedback by these protostars is still weak for accretion rates higher than 0.1 M$_{\odot}$/yr \citep{2013A&A...558A..59S,2013ApJ...778..178H}.  

The outline of this article is as follows. In section 2, we describe our numerical methodology and give a brief overview of chemical network. We present our main results in section 3 and confer our conclusions in section 4.

\section{Computational methods}

We employed the open source  code ENZO \footnote{http://enzo-project.org/, changeset:48de94f882d8} to perform three dimensional cosmological simulations following the collapse in massive primordial halos \citep{2014ApJS..211...19B}. ENZO is an adaptive mesh refinement, parallel, Eulerian code. It makes use of the piece-wise parabolic method to solve the hydrodynamical equations. The dark matter dynamics is solved by using the particle-mesh technique.

Our simulations are started with cosmological initial conditions at $\rm z=100$. We first run simulations with a uniform grid resolution of $\rm 128^3$ cells and select the most massive DM halos in a computational domain with comoving size of 1 $\rm Mpc/h$. The simulations are commenced with two additional nested refinement levels each with a resolution of $\rm 128^3$ cells in addition to a top grid resolution of $\rm 128^3$ cells and are centered on the most massive halo. We employed 5767168 particles to solve the dark matter dynamics. To follow the collapse of a halo additional 18 levels of refinement are applied during the course of simulations with a fixed Jeans resolution of 32 cells. We further make use of sink particles to follow the accretion for 10,000 years after the formation of the first sink. Detailed discussions about the sinks particles and the simulation setup can be found in our previous studies \citep{2013MNRAS.433.1607L,2013MNRAS.436.2989L,2013MNRAS.430..588L,2014MNRAS.440.1551L}.

To follow the evolution of chemical species, the rate equations of the following species $\rm H$, $\rm H^{+}$, $\rm He$, $\rm He^{+}$,~$\rm He^{++}$, $\rm e^{-}$,~$\rm H^{-}$,~$\rm H_{2}$,~$\rm H_{2}^{+}$ are solved in our cosmological simulations. We used publicly available KROME package\footnote{Webpage KROME: www.kromepackage.org} \citep{2014MNRAS.439.2386G} to compute the evolution of chemical and thermal processes. The chemical network and the microphysics employed here are the same as in \cite{2014arXiv1404.5773L} , including the self-shielding model from \cite{2011MNRAS.418..838W}. We have not included HD chemistry in our simulation as HD is very fragile molecule and even weaker flux such as $\rm J_{21}=0.1$ is found to be sufficient to dissociate it. The same was observed in simulations of \cite{2012MNRAS.426.1159S}. 

In this study, simulations are performed for two distinct halos by varying the strength of UV flux below the Lyman limit. The properties of simulated halos and the strength of UV flux in units of $\rm J_{21}$ are listed in table \ref{table1}.

\begin{table*}
\begin{center}
\caption{Properties of the simulated halos for $\rm J_{21}$ are listed here.}
\begin{tabular}{ccccccc}
\hline
\hline

Model	& $\rm J_{21}$ & Mass	 & Redshift    & spin parameter & Sink particle  masses \\

No &in units of J$_{21}$  & $\rm M_{\odot} $   &z  &  $\lambda$ & $\rm M_{\odot} $  \\
\hline                                                          \\
A  &  &  &   &0.034 & \\
   &10 & $\rm 4.3 \times 10^{7}$    & 10.98   & &1461, 231\\
   &100 & $\rm 5.45 \times 10^{7}$  & 10.63   & &7337\\
   &500 & $\rm 5.47 \times 10^{7}$   & 10.60    & &22739\\

B   &  &  &    &0.03 & \\
   &10 & $\rm 2.9 \times 10^{7}$    &11.73    & &623\\
   &100 & $\rm 3.2 \times 10^{7}$   &11.38   & &2592\\
   &500 & $\rm 3.2 \times 10^{7}$   &11.20    & & 24785\\
   
\hline
\end{tabular}
\label{table1}
\end{center}
\end{table*}

\begin{figure*}
 \hspace{-4.0cm}
\centering
\begin{tabular}{c c}
\begin{minipage}{6cm}
\includegraphics[scale=0.4]{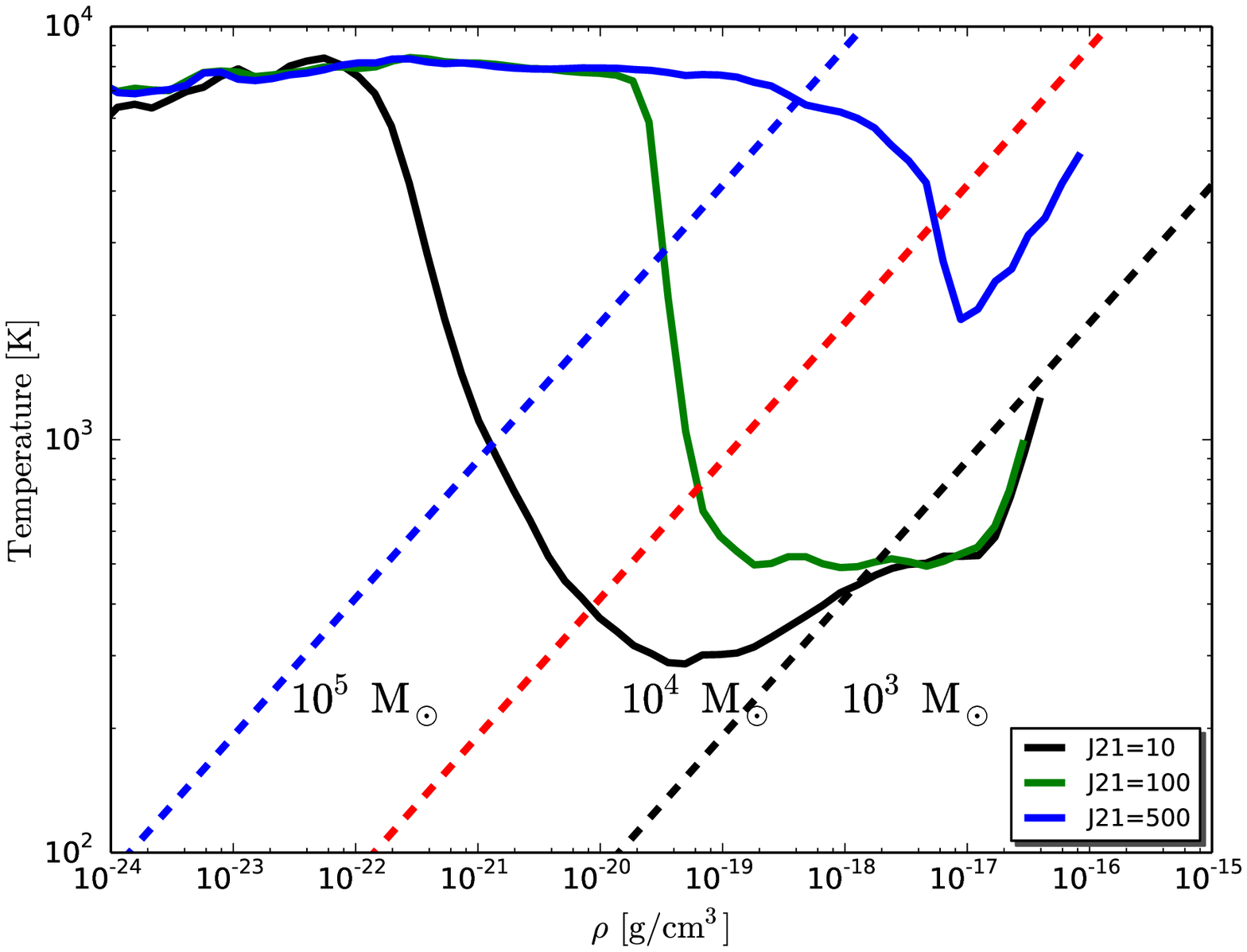}
\end{minipage} &
\hspace{2cm}
\begin{minipage}{6cm}
\includegraphics[scale=0.4]{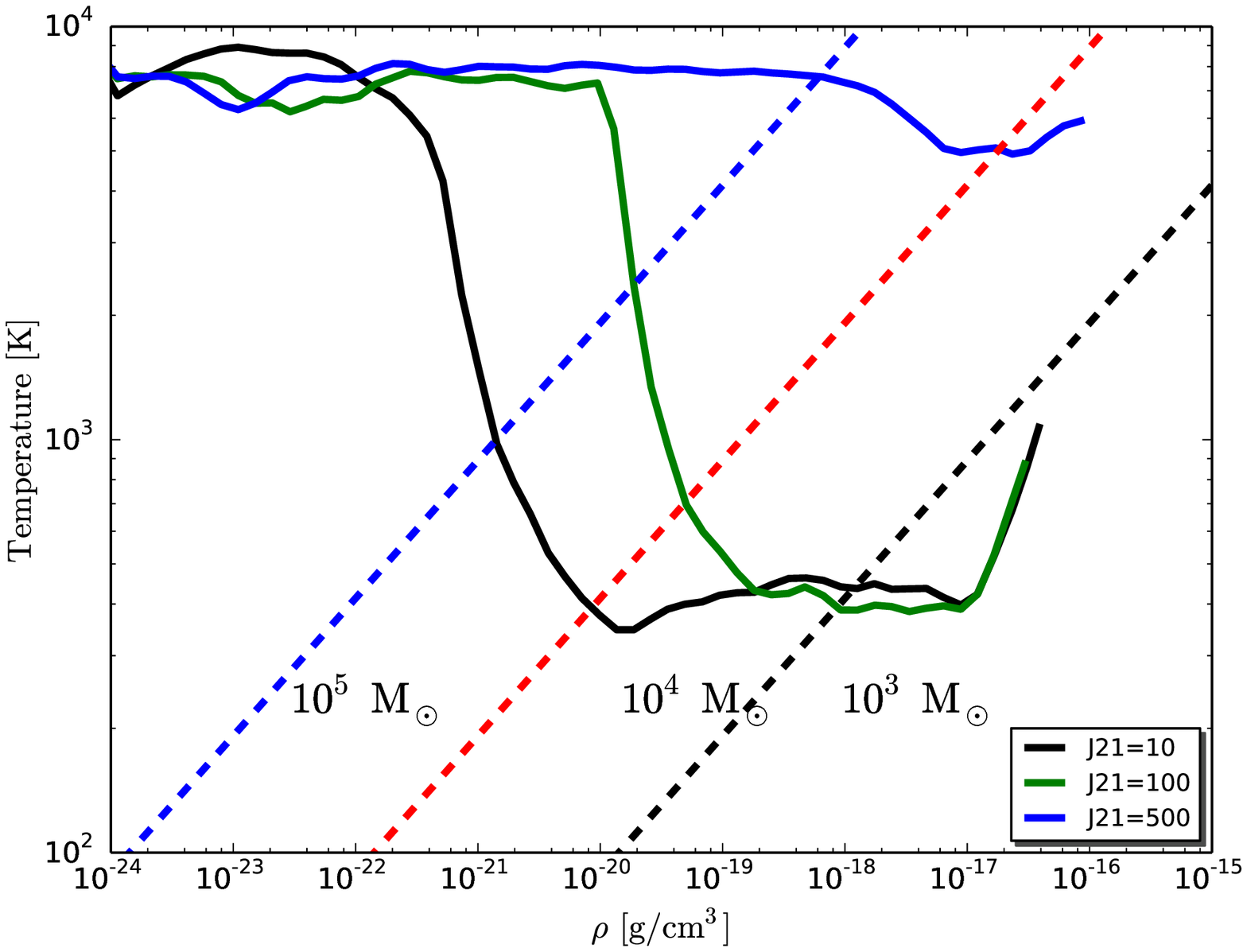}
\end{minipage}
\end{tabular}
\caption{Averaged values of temperature are plotted against density at the end point of our simulations for various strengths of $\rm J_{21}$. The left panel shows halo A and the right panel stands for halo B. The dashed lines show the expected thermal Jeans mass at these scales.}
\label{fig1}
\end{figure*}

\begin{figure*}
\centering
\begin{tabular}{c}
\begin{minipage}{6cm}
\hspace{-5cm}
\includegraphics[scale=0.3]{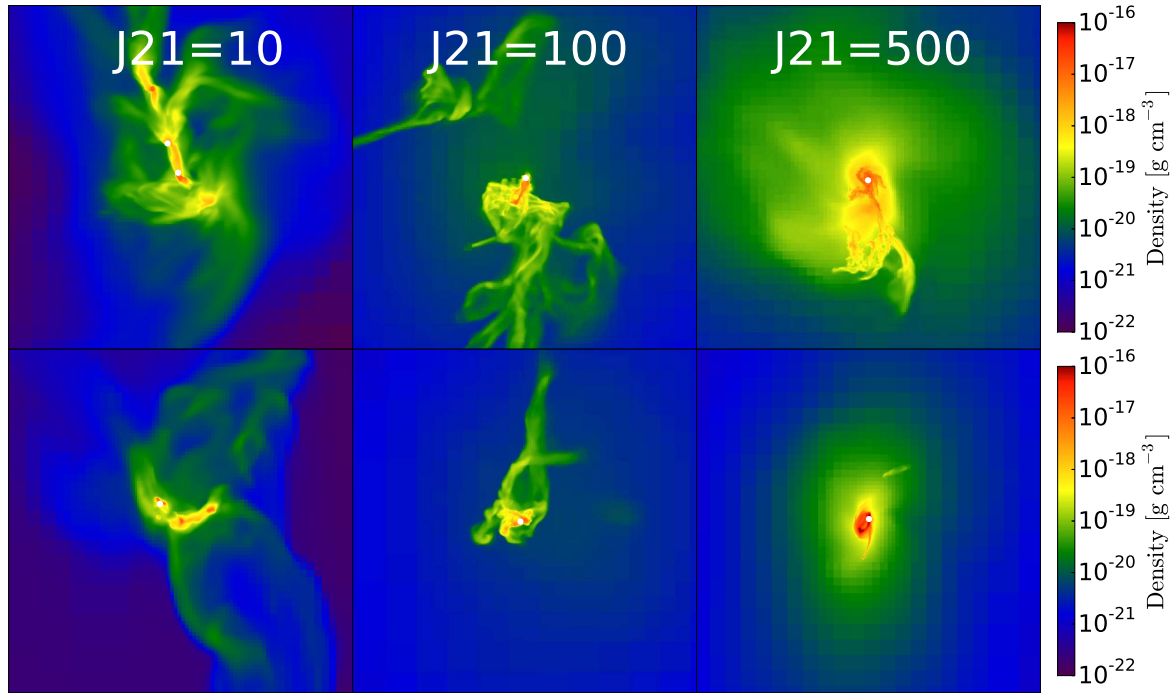}
\end{minipage} 
\end{tabular}
\caption{The final state of our simulation is represented by the average density along the line of sight for different radiation backgrounds. The top panel represents halo A and the bottom panel halo B. The overplotted white points depict sink particles. The masses of the sink particles are listed in table \ref{table1}.}
\label{fig2}
\end{figure*}

\begin{figure*}
\centering
\begin{tabular}{c}
\begin{minipage}{6cm}
\hspace{-5cm}
\includegraphics[scale=0.3]{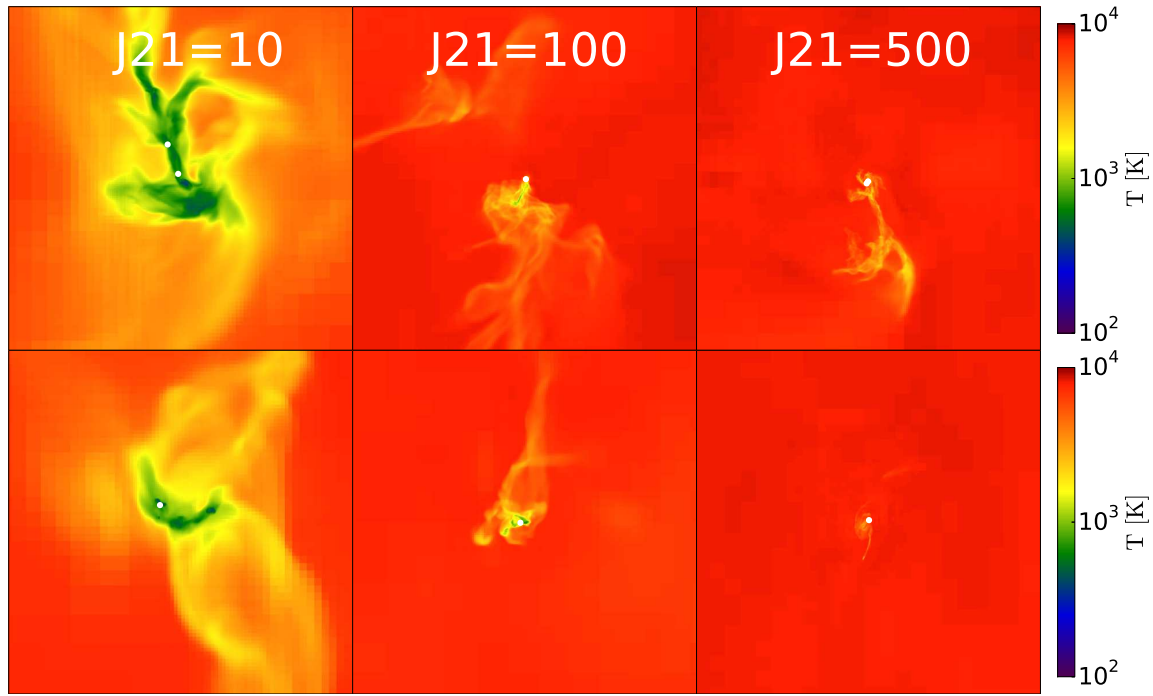}
\end{minipage}
\end{tabular}
\caption{ Temperature is shown here corresponding to the figure \ref{fig4}.}
\label{fig3}
\end{figure*}

\begin{figure*}
\centering
\begin{tabular}{c}
\begin{minipage}{6cm}
\hspace{-5cm}
\includegraphics[scale=0.3]{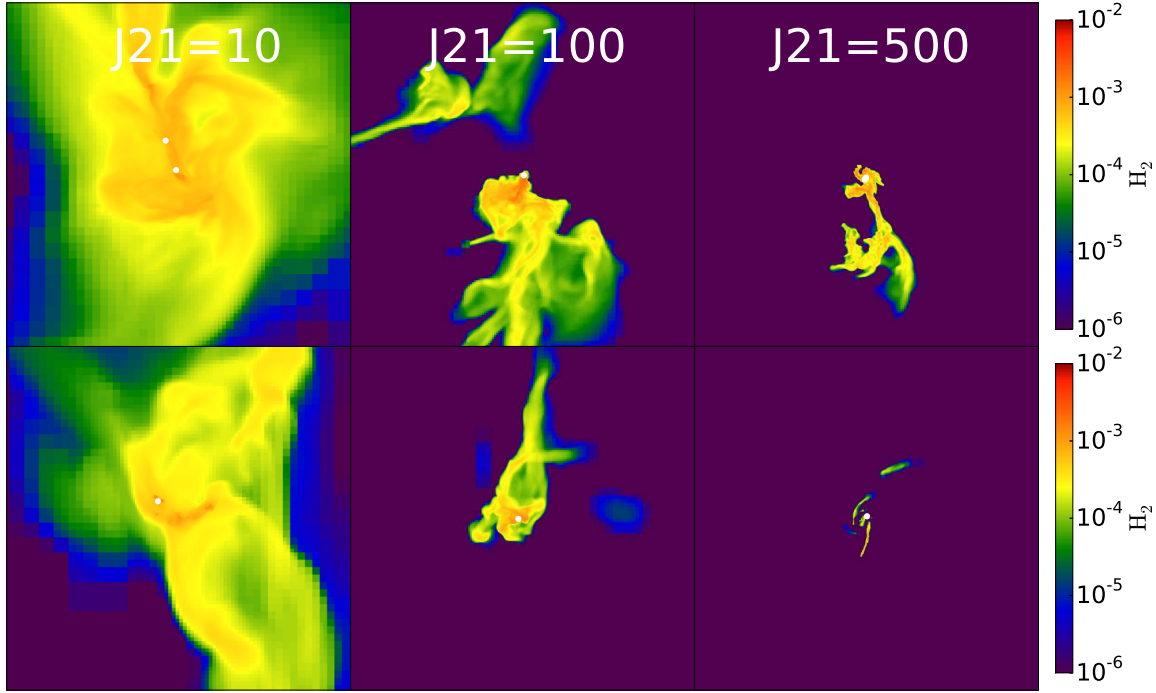}
\end{minipage}
\end{tabular}
\caption{ The abundance of molecular hydrogen is depicted here same as figure \ref{fig4}.}
\label{fig4}
\end{figure*}

\begin{figure*}
\hspace{-6.0cm}
\centering
\begin{tabular}{c}
\begin{minipage}{7cm}
\includegraphics[scale=0.66]{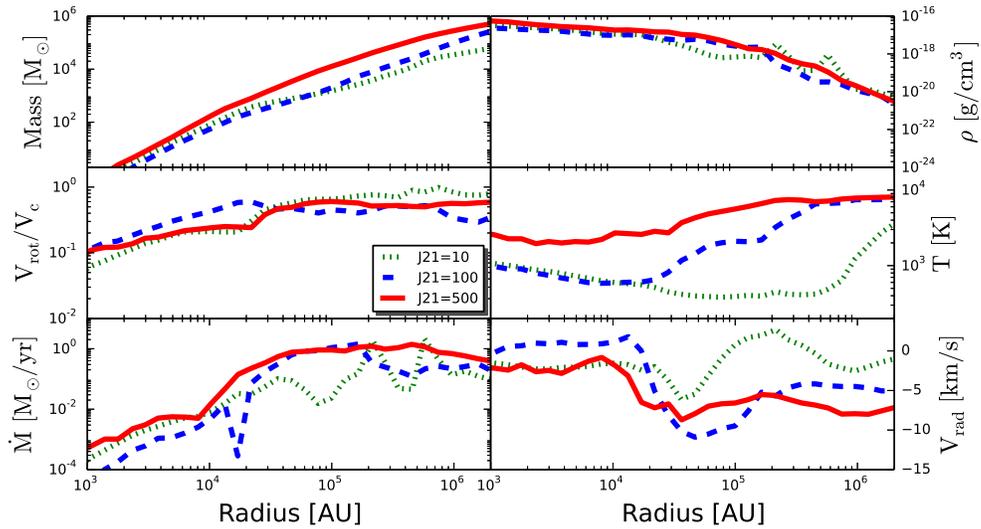}
\end{minipage}
\end{tabular}
\caption{Radially averaged spherically binned profiles are shown here for halo A at the end point of our simulations. Each line style represents a different strength of the UV flux as shown in the legend. The top panels show the mass and density, the middle panels show the ratio $\rm v_{rot}/v_{c}$ and the temperature, and the bottom panels show the mass accretion rates and infall velocities.}
\label{fig5}
\end{figure*}

\begin{figure*}
\hspace{-6.0cm}
\centering
\begin{tabular}{c}
\begin{minipage}{7cm}
\includegraphics[scale=0.66]{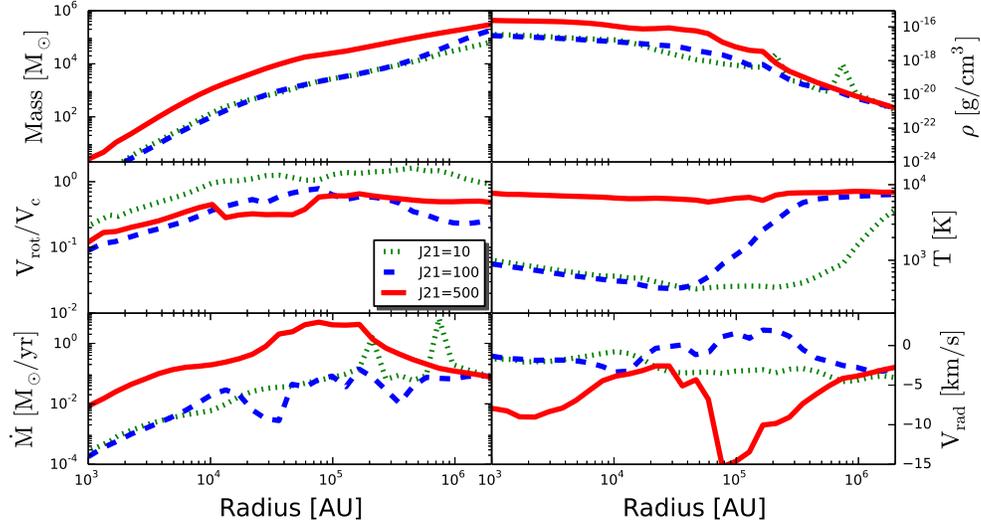}
\end{minipage}
\end{tabular}
\caption{ Same as figure \ref{fig5} but for halo B.}
\label{fig6}
\end{figure*}

\begin{figure*}
 \hspace{-4.0cm}
\centering
\begin{tabular}{c c}
\begin{minipage}{6cm}
\includegraphics[scale=0.4]{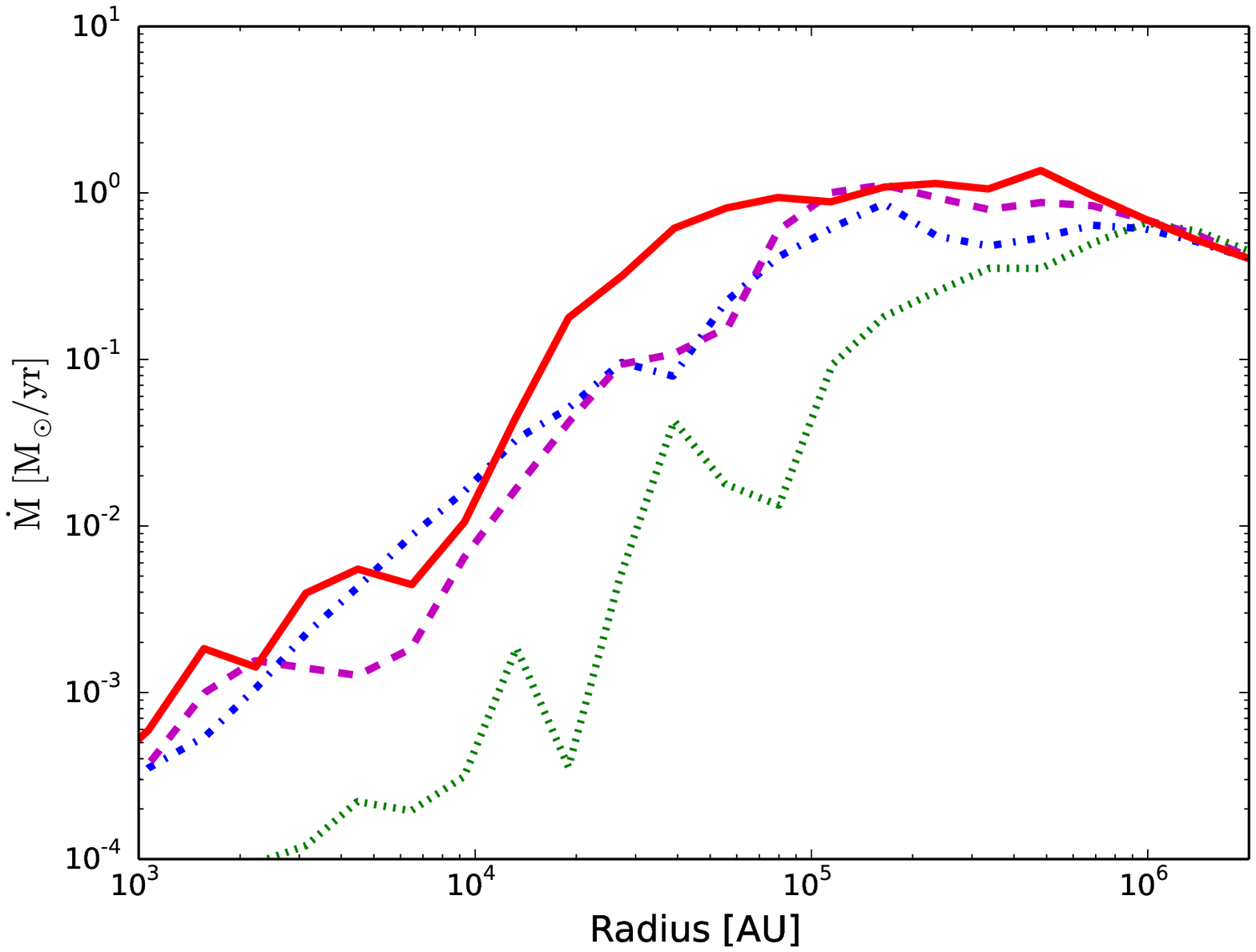}
\end{minipage} &
\hspace{2cm}
\begin{minipage}{6cm}
\includegraphics[scale=0.4]{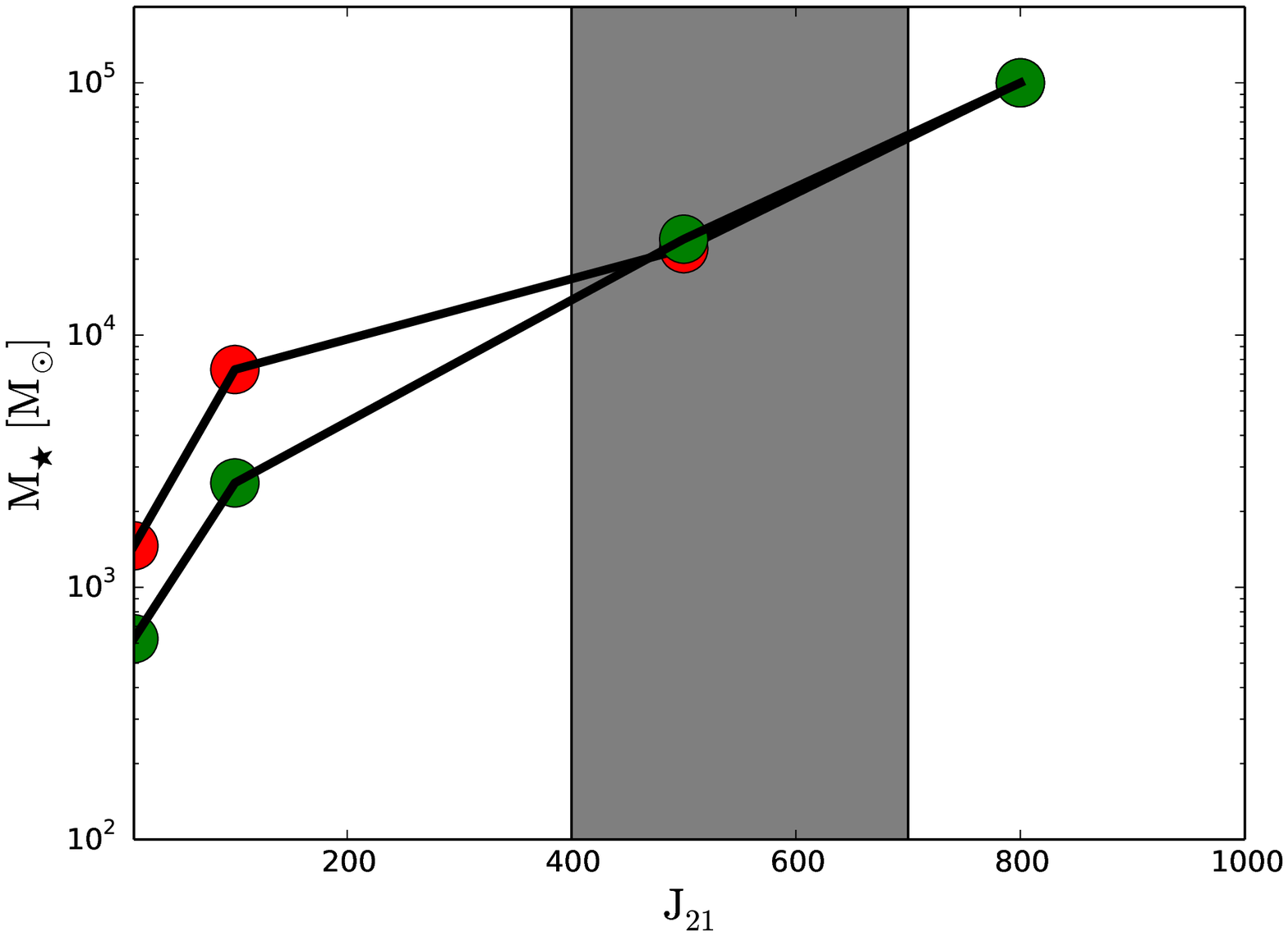}
\end{minipage}
\end{tabular}
\caption{The left panel shows the time evolution of the mass accretion rates against radius. The green line shows mass accretion rates at the formation of sink while the red line depicts evolution for 10,000 years for $\rm J_{21}=500$. The right panel shows the expected stellar masses for various strengths of the background UV flux below the Lyman limit. The dashed vertical shaded region represents the range of critical values. The blue and red spheres represent halo A and B. The last two data points are taken from Latif et al 2013e and just indicate the expected masses for $\rm J_{21}$ higher than critical values.}
\label{fig7}
\end{figure*}

\section{Main Results}

In total, we have performed six cosmological simulations for two different halos of a few times $\rm 10^{7}~M_{\odot}$ and varied the strength of $\rm J_{21}$ (i.e. 10, 100 and 500). Our results show that initially the gas is heated up to $\rm 10^4$ K in the presence of the background UV flux and then cools by Lyman alpha cooling. Depending on the strength of the UV flux, $\rm H_{2}$ formation takes place and cooling due to the molecular hydrogen becomes effective. For $\rm J_{21}=$ 10, 100, the central temperature in the halo is about 1000 K while for $\rm J_{21}=500$ the temperature in the core is a few thousand K as shown in figure \ref{fig1}. It is also found that for a weaker flux (i.e. $\rm J_{21}=$ 10), $\rm H_{2}$ cooling kicks in at densities around $\rm 10^{-23}~g/cm^{3}$ while for $\rm J_{21}=$ 100 it becomes effective around $\rm 10^{-20}~g/cm^{3}$. For the strongest flux, it starts at even higher densities of about $\rm 10^{-18}~g/cm^{3}$.

The density structure for various strengths of the UV field is shown in figure \ref{fig2}. It can be seen that for $\rm J_{21}=$ 10 the gas in the halos becomes more clumpy and indicates the possibility to form multiple clumps. These clumps are formed by locally induced $\rm H_{2}$ cooling which lowers the gas temperature down to a few hundred K as shown in figures \ref{fig3} and \ref{fig4}. For $\rm J_{21}=$ 10, the clumps with no sinks are gravitational unbound and have masses of the order of few solar masses. For $\rm J_{21}=$ 100, the halos have less structure and the number of cold clumps is reduced. This is due to the fact that the overall temperature in the halo is higher particularly in the outskirts which suppresses fragmentation. Similarly, the fraction of molecular hydrogen is significantly reduced in the surrounding of the halo. For the strongest flux ($\rm J_{21}=$ 500), there is no sign of fragmentation as the central temperature in the halos is a few thousand K and in one halo later approaches an isothermal state due to the temperature dependence of the $\rm H_{2}$ collisional dissociation rate \citep{1996ApJ...461..265M}. The fraction of molecular hydrogen is significantly lower and is limited to only core of the halos.

To further quantify the masses of the clumps formed in our simulated halos, we employed sink particles which represent the protostars in our simulations and followed the evolution for many free-fall times. The masses of the sinks are listed in table \ref{table1}. It is found that for  $\rm J_{21}=$ 10, the masses of the sinks are a few hundred to about 1500 M$_{\odot}$. In halo A two sinks are formed. For $\rm J_{21}=$ 100, only one massive sink is formed in each halo and their masses are 7400 and 2500 M$_{\odot}$. The masses of the sinks are above 22, 000 solar masses for the strongest flux case and a single sink is formed per halo. Overall, our results suggest that fragmentation in the halos is reduced by increasing the strength of background UV flux according to the theoretical expectations. For $\rm J_{21}=$ 10 and 100, although the central temperatures are quite similar the masses are higher by a factor of a few for $\rm J_{21}=$ 100. This is because of the warmer gas in the surroundings of the halo which leads to higher accretion rates and consequently higher sink masses at the end of our simulations. It is found that the masses of the sinks in halo B are about a factor of two lower compared to halo A. Overall, the average density in halo B is lower and has a higher rotational support. This leads to lower accretion rates in halo B and consequently lower sink masses.

The physical properties of the halos are depicted in figures \ref{fig5} and \ref{fig6}. The maximum density in the halos is $\rm 10^{-16}~g/cm^{3}$. The density profile follow an R$^{-2}$ behavior at larger radii for almost all fluxes and becomes flat in the core of the halo corresponding to the Jeans length. The deviations from this behavior are observed for weaker values of $\rm J_{21}$ due to the additional substructure inside the halo. The difference between weaker and strongest flux are more prominent in halo B as it collapses almost isothermally for $\rm J_{21}=500$. The overall density in halo B is lower by a factor of a few compared to the halo A around 10 pc. For $\rm J_{21}=10$, cooling due to $\rm H_{2}$ becomes effective in the central 10 pc of the halo. On the other hand for  $\rm J_{21}=100$ it is limited to the central pc of the halo while for $\rm J_{21}=500$ cooling due to the molecular hydrogen is not very effective due to its lower abundance. 

The impact of different thermal evolutions is also reflected in the infall velocities of the halos. Higher temperatures lead to higher infall velocities and also indicate lower molecular hydrogen fractions in those halos. Similarly lower infall velocities are observed for the weaker fluxes indicating that the halo is in the molecular cooling phase. Halo B has lower radial infall velocities compared to halo A. Large accretion rates of about 0.1~M$_{\odot}$/yr are observed. The decline in accretion rates towards the center is due to the increasing pressure support within the Jeans length. It is further found that the accretion rates are slightly higher for stronger fluxes and are lower in halo B compared to halo A. The enclosed mass in the halos is higher for stronger fluxes. This is due to the higher infall velocities and higher accretion rates for these cases. It is also found that halo B has higher rotational support. We further noted that the accretions rates increased during the course of simulations and are shown for a representative case in figure \ref{fig7}. 

We show the expected stellar masses for various values of $\rm J_{21}$ in figure \ref{fig7}. It is observed that the stars formed in the presence of UV feedback are very massive typically above one thousand solar masses. This figure shows an increase in the stellar masses with the strength of the UV flux, i.e. the higher the UV flux the more massive the star. Fragmentation may occur at larger scales but may not be able to prevent the formation of massive stars. We expect that these massive to supermassive stars are the potential candidates for the formation of intermediate to supermassive black holes.
 
\section{Discussion and conclusions}

In this article, we explored fragmentation in massive primordial halos of a few times $\rm 10^{7}$ M$_{\odot}$ irradiated by different UV fluxes below the Lyman limit. To achieve this goal, we performed high resolution cosmological simulations for two distinct halos collapsing at $\rm z>10$ and varied the strength of the UV flux from 10-500 in units of $\rm J_{21}$. We exploited the adaptive mesh refinement technique to follow the collapse of the halo by employing additional 18 dynamical refinement levels during the course of simulations which yields an effective resolution of about 100 AU. To further follow the evolution for longer time scales, sink particles representing protostars were employed. A fixed Jeans resolution of 32 cells were mandated during the entire course of simulations.

Our findings show that the formation of $\rm H_{2}$ gets delayed and no strong fragmentation occurs in the halos illuminated by the UV flux of  10-500 in units of $\rm J_{21}$. For $\rm J_{21}=10$, in our simulated halos binary or multiple systems may likely form but for higher values of $\rm J_{21}$ only one single object is formed. At the end of our simulations, massive protostars of a few 100-10, 000 solar masses are formed and large accretion rates of about $0.1-0.01$~M$_{\odot}$/yr are observed. These relatively high accretion rates distinguish them from normal star formation mode in minihalos. These massive stars formed in the early universe are the potential candidates for the formation of intermediate mass black holes.

$\rm H_{2}$ cooling reaches local thermal equilibrium at densities higher than $\rm 10^4~cm^{-3}$ and the cooling rate scales as density while compressional heating varies with $\rho^{3/2}$. So the expected polytropic index stiffens (i.e. $\gamma$ $\rm >$1) and suppresses fragmentation \citep{2000ApJ...538..115S}. Of course, the possibility of fragmentation at higher density cannot be completely ruled out and low mass stars may form in addition to the most massive object.


\cite{2012MNRAS.426.1159S} performed similar simulations for a halo of similar mass irradiated by the background UV flux emitted by population III stars. They also found that the gas self-shields from the UV flux and a dense core of $10^{4}$~M$_{\odot}$ develops. They further used sink particles to follow the simulations for longer times and found that a massive sink of about 1000 solar masses is formed. Their study is comparable to $\rm J_{21}=10$ in our simulations due to the background UV flux emitted by two different stellar populations. They have employed radiation spectrum with $\rm T_{*}=10^5$~K in their calculations which mainly dissociates the molecular hydrogen via the Solomon process and $\rm H_{2}$ self-shielding is more important in contrast to our study. 

In particular, we compare the accretion rates at the same times in both simulations and our results appear rather consistent. We note however that there is no one-to-one correspondence, as the thermal evolution does not precisely match and our halo is somewhat warmer on larger scales. In addition, there are of course fluctuations from halo to halo, so some variation between the halos is expected, as we also see here when comparing halos A and B. 

To further explore the possible fragmentation at higher densities, we in total added 27 refinement levels which yield resolution down to sub AU scales (0.25~AU) and the peak densities of $\rm 10^{-11}~g/cm^{3}$. We however do not see any indication of further fragmentation for these runs. Such verification further confirms that even if fragmentation occurs at later stages the mass of central object would not change significantly. Our results for $\rm J_{21}=10$ are in agreement with \cite{2014ApJ...781...60H} who simulated about 100 minihalos and found that even massive stars of thousand solar masses can form in the presence of similar mass accretion rates that found in this study.

We also note that our simulations do not include the ionizing UV feedback by the protostar. As long as the accretion rates remain higher than $\rm 0.1$~M$_{\odot}$/yr, no strong feedback is expected \citep{2012ApJ...756...93H,2013ApJ...778..178H,2013A&A...558A..59S}. Ionizing feedback may kick in once the accretion rates drop more significantly, potentially limiting the growth of the resulting objects. 

We also assumed that halos are metal free and given the patchy distribution of the metal such halos may exist down to z=6 \citep{2009ApJ...700.1672T}. However, if metal enrichment takes places in these halos it will lead to fragmentation and most likely star cluster of low mass stars \citep{2014MNRAS.440L..76S,2014arXiv1406.0346P}. Even a small amount of dust, i.e. $\rm 10^{-5}~ Z/Z_{\odot}$ is sufficient to boost the H$_{2}$ fraction and induce fragmentation \citep{2009A&A...496..365C}. In fact, recent observations suggest the potential presence of intermediate mass black holes in globular clusters \citep{2014arXiv1404.2781K}. 

From the results given here, we deduce that massive to supermassive stars may be more common than previously expected, and can also form for radiative backgrounds $\rm J_{21}<J_{crit}$. Based on the recent estimates of \cite{2014arXiv1405.6743D} for the expected density of halos for various UV field strengths, the expected fraction of halos exposed to $\rm J_{21}<J_{crit}$ is few orders of magnitude higher. We expect such stars to produce a considerable amount of UV feedback once they are on the main sequence, thus considerably contributing to the ionization of their environment, and perhaps also to the epoch of reionization. At the end of their lives, they may collapse to an intermediate mass black hole, and the X-rays released during their formation may contribute to establish a global background \citep{2013MNRAS.433.1556Y}.

\section*{Acknowledgments}
The simulations described in this work were performed using the Enzo code, developed by the Laboratory for Computational Astrophysics at the University of California in San Diego (http://lca.ucsd.edu). We acknowledge research funding by Deutsche Forschungsgemeinschaft (DFG) under grant SFB $\rm 963/1$ (project A12) and computing time from HLRN under project nip00029. DRGS and SB thank the DFG for funding via the Schwerpunktprogram SPP 1573 ``Physics of the Interstellar Medium'' (grant SCHL $\rm 1964/1-1$). The simulation results are analyzed using the visualization toolkit for astrophysical data YT \citep{2011ApJS..192....9T}.


\end{document}